\documentclass[letterpaper,english,aps,prl,twocolumn,floatfix,showpacs,amsfonts,amssymb,superscriptaddress]{revtex4}

\usepackage[T1]{fontenc}
\usepackage[latin1]{inputenc}
\usepackage{graphics}
\usepackage{amssymb}
\usepackage{overpic}

\newcommand{\beq}{\begin{equation}}
\newcommand{\eeq}{\end{equation}}
\newcommand{\bea}{\begin{eqnarray}}
\newcommand{\eea}{\end{eqnarray}}

\begin{document}

%\title{First order phase transition and spinon deconfinement in the $J_{1}$-$J_{2}$ Heisenberg model}
\title{Magnetic quantum phase transitions of the antiferromagnetic $J_{1}$-$J_{2}$ Heisenberg model}

\author{T.~P.~Cysne}

\affiliation{Instituto de F\' isica, Universidade Federal do
Rio de Janeiro, Caixa Postal 68528, Brasil}

\author{M.~B.~Silva~Neto}

\affiliation{Instituto de F\' isica, Universidade Federal do
Rio de Janeiro, Caixa Postal 68528, Brasil}

\received{...}

\begin{abstract}

We obtain the complete phase diagram of the antiferromagnetic $J_{1}$-$J_{2}$ 
model, $0\leq \alpha = J_2/J1 \leq 1$, within the framework of the $O(N)$ nonlinear 
sigma model. We find two magnetically ordered phases, one with N\' eel order, for 
$\alpha \leq 0.4$, and another with collinear order, for $\alpha\geq 0.6$, separated by 
a nonmagnetic region, for $0.4\leq \alpha \leq 0.6$, where a gapped spin liquid 
is found. The transition at $\alpha=0.4$ is of the second order while the one at 
$\alpha=0.6$ is of the first order and the spin gaps cross at $\alpha=0.5$. Our 
results are exact at $N\rightarrow\infty$ and agree with numerical results 
from different methods.

\end{abstract}


\maketitle

Quantum phase transitions (QPTs) occur when the ground state 
properties of a certain physical system undergo dramatic 
changes as one, or more, internal or external, parameters are 
varied \cite{Book-Sashdev}. Examples include, but are not 
restricted to, magnetic phase transitions between two distinct 
magnetic ground states or between a magnetic state and a nonmagnetic one,
driven for example by an applied field, pressure, or the coupling 
to oder degrees of freedom. QPTs are usually labelled according 
to the behaviour of some order parameter (OP) close to the 
quantum critical point (QCP) \cite{Goldenfeld}, and are said to be of the second 
order (2nd order) when the OP vanishes continuously as the 
QCP is approached, or of the first order (1st order) when the 
OP has a finite value near the QCP and jumps discontinuously 
to zero above it. Furthermore, knowledge of the range of 
the interactions, symmetries of the Hamiltonian and 
dimension of the OP, allow us to classify QPTs into 
universality classes \cite{Goldenfeld}, and help us to wirte down 
a Landau-Ginzburg free energy (LGFE) to describe such 
phase transitions (PTs). Typically, LGFEs up to the 4th power 
of the OP are enough to describe a 2nd order PT, while 
LGFEs up to the 6th power of the OP are 
necessary to describe a 1st order PT.

The $O(N)$ quantum nonlinear sigma model (NLSM) has long
been acknowledged to be a very convenient framework to
describe 2nd order magnetic PTs in spin systems, such 
as, for example, the antiferromagnetic (AF) Heisenberg 
Hamiltonian, in two dimensions, with nearest-neighbour
interactions on a square lattice \cite{NLSM-Heisenberg}. 
Here the QPT occurs between a N\' eel ordered magnetic 
ground state, where the OP is the sublattice magnetization, 
$\sigma\neq 0$, and a nonmagnetic state ($\sigma=0$) 
with a finite spin gap, $\Delta\neq 0$, as the OP at zero 
temperature. Such transition is driven by quantum 
fluctuations set by some coupling constant, $g$, 
and is of the 2nd order, as both $\sigma$ and $\Delta$ 
vanish continuously at the QCP, $g_c$. Despite being 
nonlinear, at the mean field level ($N\rightarrow\infty$) 
the model is quadratic, exactly solvable, and produces 
the usual mean field values for the critical exponents of 
the Heisenberg universality class, 
$\sigma\propto (g_c-g)^\beta$, for the ordered regime
($g<g_c$), with $\beta=1/2$, and $\Delta\propto (g-g_c)^{\nu}$,
for the nonmagnetic phase ($g>g_c$), with $\nu=1$
\cite{Chubukov-Sachdev-Ye}.

First order PTs in spin systems occur whenever two 
magnetic phases cannot be continuously connected to 
one another by some order parameter. This is what 
happens, for example, already at the classical level, 
between the N\' eel- and collinear-type ordering phases 
of the $J_1-J_2$ model, at the border $\alpha = 0.5$. 
When quantum fluctuations are taken into account, 
a nonmagnetic region opens up around $\alpha=0.5$ 
\cite{Review-J1-J2} and a gapped spin liquid phase 
is found for $0.4\leq\alpha\leq 0.6$ \cite{Spin-Gap}. 
Although the precise nature 
of the nonmagnetic state is still under debate (typical 
candidates range from dimer to plaquette or VBS 
phases) the nature of the transition at $\alpha=0.4$ 
is agreed to be of the 2nd order by either numerical 
and theoretical methods, like for example the NLSM
\cite{NLSM-J1-J2}. For the 
transition at $\alpha=0.6$, different numerical techniques, 
including series expansion \cite{Series-Expansion}, 
quantum Monte Carlo \cite{QMC}, 
exact diagonalization \cite{Exact-Diag}, and 
DMRG \cite{DMRG}, strongly indicate 
it to be of the 1st order \cite{First-Order-J1-J2}, but 
from the theoretical point of view no conclusive 
statement has yet been presented. More importantly, 
this poses serious questions on the applicability of 
the NLSM to describe a 1st order PT in frustrated 
magnetic systems \cite{Senthil-Science}, specially since 
no unusual powers of the OP are to be expected. 
 
In this work we derive and apply the $O(N)$ NLSM formalism 
for the $J_1-J_2$ Heisenberg model, for the whole range 
of parameters, $0\leq \alpha \leq 1$. Up to the classical border, 
$0\leq\alpha\leq 0.5$, the model describes smooth
fluctuations of the staggered order parameter on top of a
N\' eel ordered ground state and possesses a 2nd order
phase transition, at $\alpha=0.4$, driven by quantum fluctuations, 
towards a nonmagnetic, gapped spin liquid phase. 
Beyond the classical border, $0.5\leq\alpha\leq 1$, the 
model describes, instead, smooth fluctuations of the staggered 
order parameter on top of a collinearly ordered ground state.
Remarkably, although at the mean field ($N\rightarrow\infty$) 
level the model remains quadratic and exactly solvable, we 
show that its quantum dynamics is importantly modified by 
a term proportional to the AF order parameter, which 
causes significant changes on the behaviour of the OP 
at zero temperature. The nonmagnetic, gapped spin
liquid and collinear phases can no longer be continuously
connected and a 1st order QPT is theoretically obtained. 

The $J_{1}-J_{2}$ Heisenberg spin-Hamiltonian
is given by
\beq
\hat{H}=J_{1}\sum_{\langle i,j\rangle}{\hat{\bf S}}_{i}\cdot {\hat{\bf S}}_{j}+
J_{2}\sum_{\langle\langle i,j\rangle \rangle}{\hat{\bf S}}_{i}\cdot {\hat{\bf S}}_{j},
\label{J1-J2-Hamiltonian}
\eeq
where $J_{1}>0$ and $J_{2}>0$ are, respectively, the AF
superexchanges between nearest-neighbors, $\langle i,j\rangle$, and 
next-to-nearest neighbors, $\langle \langle i,j\rangle \rangle$, spins 
$\hat{\bf S}_i$ on a two dimensional square lattice. 
The Hamiltonian (\ref{J1-J2-Hamiltonian}) 
exhibits two types of magnetic order: N\' eel order, with 
wave vector at $q=(\pi,\pi)$, for $\alpha\leq 0.4$, and collinear order, with 
wave vectors at $q=(\pi,0)$ and/or $q=(0,\pi)$, for $\alpha\geq 0.6$ \cite{Dai}. 

%**********************************************************************************************
\begin{figure}[t]
\includegraphics[scale=0.32]{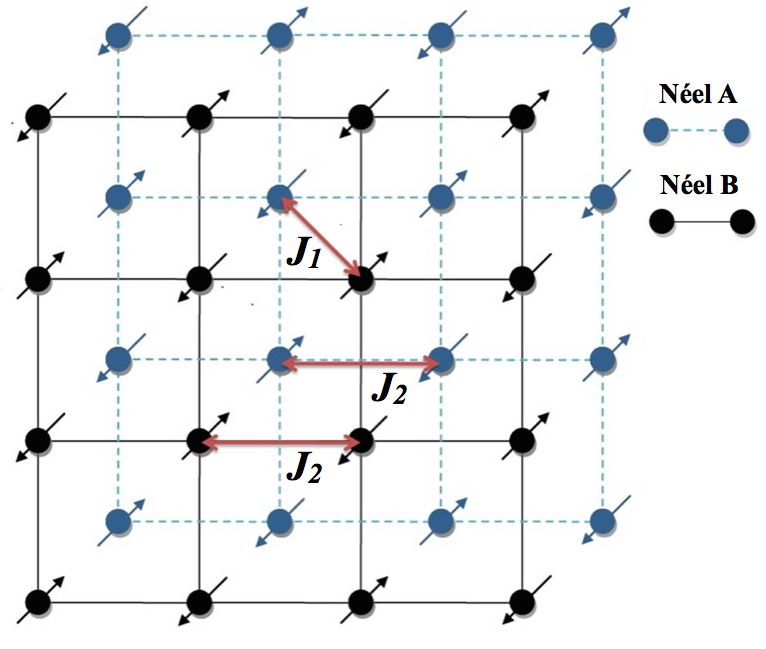}
\caption{Collinear state of the $J_1-J_2$ model for $0.6\leq\alpha\leq 1$
as two interpenetrated N\' eel structures $A$ and $B$, with nearest neighbour,
$J_1$, and next-to-nearest neighbour, $J_2$, AF interactions, upon which 
we shall build up our NLSM.}
\label{Fig-Collinear-GS}
\end{figure}
%**********************************************************************************************

For the N\' eel $(\pi,\pi)$ phase, different effective field theories, of the NLSM 
type, have been proposed \cite{NLSM-J1-J2}, and they 
all succeed in describing the 2nd order PT at $\alpha=0.4$. For the 
collinear $(\pi,0)$ and/or $(0,\pi)$ phase, instead, no such description has been provided yet,
and we shall proceed as follows: we treat the collinear magnetic state 
as a result of two interpenetrated N\' eel ordered sublattices and introduce 
an double coherent spin-state basis, with spin operators labeled by 
indices $A$ and $B$, see Fig.~\ref{Fig-Collinear-GS}. We then associate 
the spins operators in Eq. (\ref{J1-J2-Hamiltonian}) to vector 
fields $\vec{n}_A(i)$ and $\vec{n}_B(i)$ that describe long wavelength deviations
from the N\' eel state in each sublattice. As usual we parametrize the spin-1 
fields into a smooth, $\vec{m}$, and a fast and uniform, $\vec{L}$, varying components,
$\vec{n}_{A,B}=\theta_{A,B}\vec{m}_{A,B}\sqrt{1-(\bar{a}L_{A,B})^2}+\bar{a}\vec{L}_{A,B}$,
where $\theta_{A,B}(i)=+1$ for spin $\uparrow$ and $=-1$ for spin $\downarrow$, 
and $\bar{a}=a^d/S$. To satisfy $\vec{n}_{A,B}\cdot\vec{n}_{A,B}=1$ we assume that 
$\vec{m}_{A,B}\cdot\vec{m}_{A,B}\approx1$, while $\vec{m}_{A,B}\cdot\vec{L}_{A,B}\approx0$, 
$\bar{a}^2\vec{L}_{A,B}\cdot\vec{L}_{A,B}<<1$, and
$\vec{m}_A\cdot\vec{L}_B=\vec{m}_B\cdot\vec{L}_A=0$.
After integration over $\vec{L}$ the action for the smooth fields is 
\bea
S&=&\frac{\rho_S}{2} \int \left[ (\nabla\vec{m}_A)^2+(\nabla\vec{m}_B)^2
+ \frac{(\partial_{\tau}\vec{m}_A)^2+(\partial_{\tau}\vec{m}_B)^2}{c_0^2} \right. \nonumber \\
&+& \left. \gamma_0(\vec{m}_B\cdot\partial_x\partial_y\vec{m}_A+\vec{m}_A\cdot\partial_x\partial_y\vec{m}_B) 
\right.\nonumber\\
&-& \left. c_1^{-2}(\vec{m}_A\times\partial_{\tau}\vec{m}_A)\cdot(\vec{m}_B\times\partial_{\tau}\vec{m}_B) \right]  \label{s2},
\label{Collinear-NLSM}
\eea
where $\rho_S=2J_2S^2$ is the  spin stiffness in two dimensions, $c_0=\sqrt{2}Sa\sqrt{16J_2^2-J_1^2}$ and 
$c_1=\sqrt{2\frac{J_2}{J_1}}c_0$ are spin-wave velocities, and 
$\gamma_0=\frac{J_1}{J_2}$. The first line in Eq.~(\ref{Collinear-NLSM}) 
corresponds to the usual NLSM for the two 
N\' eel sub-structures of Fig.~\ref{Fig-Collinear-GS}, labelled $A$ and $B$, 
which are decoupled when $J_1=0$. For 
$J_1\neq 0$, however, two couplings arise: the first one
involves only gradient terms and produces different spin-wave
velocities along the diagonals \cite{Conceicao}; 
the second, and more important one, is a result of the coupled pressession of magnetic 
moments on the two N\' eel sub-structures and modifies importantly the dynamics of the 
problem, ultimately leading to the first order character of the phase transition at $\alpha=0.6$.

In the magnetically ordered phase we can write
$\vec{m}_{A,B}=\pi_{x,(A,B)}\hat{x}+\pi_{y,(A,B)}\hat{y}+\sigma\hat{z} $.
The $\pi$ fields are associated to the quantum fluctuations and the 
$\sigma$ field to the staggered OP. We introduce the
Lagrange multiplier
$S_{vinc}\propto\int i\lambda(\big|\vec{m}_A\big|^2-1)+ i\lambda(\big|\vec{m}_B\big|^2-1)$
and after integrating out transverse fluctuations
we end up with the partition function
$Z(\beta)=\mathcal{N'}\int D[i\lambda]D[\sigma]e^{-NS_{eff}[\lambda,\sigma]}$,
where 
\begin{eqnarray}
S_{eff}[\lambda,\sigma]=\frac{N-1}{N}{Tr} \ln \big(A(\partial)+i\lambda\mathbb{I})+\int\frac{2}{gc_0}i\lambda(\sigma^2-1)\nonumber
\end{eqnarray}
is given in terms of
\begin{eqnarray}
 A(\partial)=\left[ \begin{array}{rrrr} 
a_1(\partial) & 0 \ \ \ \ \  & a_2(\partial) & 0 \ \ \ \ \   \\ 
 0 \ \ \ \ \   & a_1(\partial) & 0 \ \ \ \ \  & a_2(\partial) \\ 
 a_2(\partial) & 0 \ \ \ \ \  & a_1(\partial) & 0 \ \ \ \ \   \\ 
 0 \ \ \ \ \    & a_2(\partial) & 0 \ \ \ \ \   & a_1(\partial) \end{array} \right],
\end{eqnarray}
with $a_1(\partial)=-c_0^2\nabla^2-\partial_{\tau}^2$,
$a_2(\partial,\sigma)=\gamma_0 c_0^2\partial_x\partial_y+\frac{\sigma^2}{2}v\partial^2_{\tau}$,
$v=c_0^2/c_1^2$, and
$g=\frac{N\hbar c_0}{\rho_S}=2\sqrt{2}a\frac{N}{S}\sqrt{1-\frac{1}{16}\Big(\frac{J_1}{J_2}\Big)^2}$
determines the strength of the coupling between quantum fluctuations (set by $1/S$) and
frustration (set by $J_1/J_2$). 

%**********************************************************************************************
\begin{figure}[t]
\includegraphics[scale=0.4]{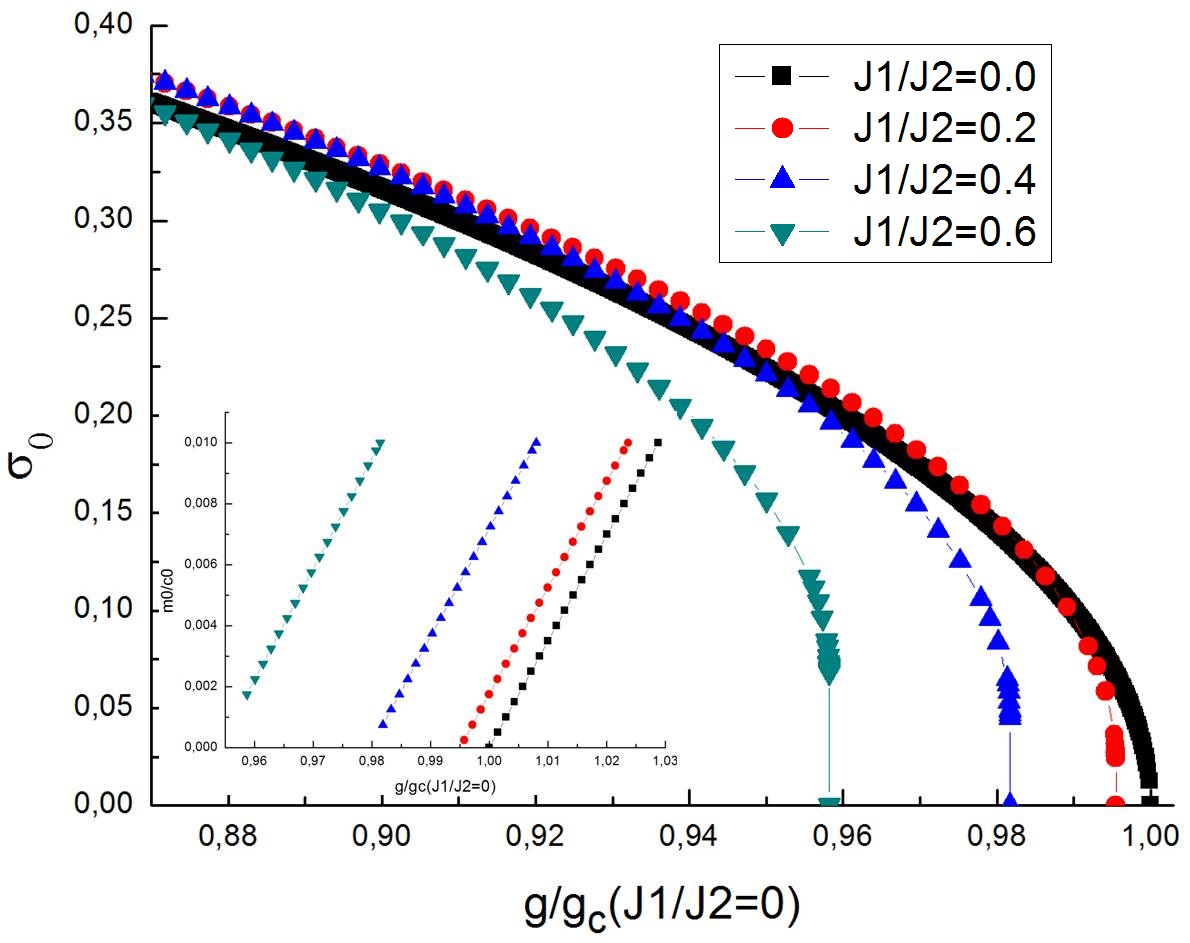}
\caption{Solutions to Eqs. (\ref{Eqs-SP}) for the OP $\sigma_0$ and the spin gap 
$m_0/c_0$ (inset), at $T=0$, as a function of $g$, for different small ratios of $J_1/J_2$. 
For $J_1=0$ (black squares),  the OP and the spin gap vanish continuously 
at $g_c$, and the PT is of the 2nd order (Heisenberg model). For $J_1\neq 0$ 
(red, blue and green symbols), however,  the OP and the spin gap jump 
discontinuously to zero at $g_c$, indicating a 1st order PT.} 
\label{Fig-Magnetization}
\end{figure}
%**********************************************************************************************

In the limit  $N\rightarrow \infty$ we look for solutions of the type
$\sigma(\vec{x},\tau)=\sigma_0$ and $i\lambda(\vec{x},\tau)=m_0^2$,
where $\sigma_0$ and $m_0^2$ are given by
$\frac{\partial S_{eff}[\lambda,\sigma]}{\partial \sigma}\Bigg]_{\sigma=\sigma_0}=0$,
and
$\frac{\partial S_{eff}[\lambda,\sigma]}{\partial i\lambda}\Bigg]_{i\lambda=m_0^2}=0$.
The saddle point equations in the large $N$ limit and for the magnetically order
phase, where $\sigma_0\neq 0$, then become 
\begin{eqnarray}
\left\{ 
\begin{array}{c}
\sigma_0^2=f(m_0,\sigma_0) 
 \\  (\frac{m_0}{c_0})^2=h(m_0,\sigma_0).
\end{array}
\right.
\label{Eqs-SP}
\end{eqnarray}
We are interested in the quantum phase transition in which case
$\frac{1}{\beta\hbar}\sum_{\omega_n} \rightarrow \int\frac{d\omega}{2\pi}$,
and thus \cite{Next}
\begin{eqnarray}
f(m_0,\sigma_0)&=&1-g\int\frac{d^3k}{(2\pi)^3}\Big[ G^+_{\vec{k}}(m_0,\sigma_0)+G^-_{\vec{k}}(m_0,\sigma_0) \Big], \nonumber \\
h(m_0,\sigma_0)&=&\frac{gv}{2}\int\frac{d^3k}{(2\pi)^3}k_z^2\Big[G^-_{\vec{k}}(m_0,\sigma_0)-G^+_{\vec{k}}(m_0,\sigma_0) \Big], \nonumber
\end{eqnarray}
where $k_z=\omega/c_0$. The Green's functions are
$G^{\pm}_{\vec{k}}(m_0,\sigma_0)=\frac{1}{D_{\vec{k}}^{\pm}(\sigma_0)+\big(\frac{m_0}{c_0}\big)^2}$, 
where we have defined $D_{\vec{k}}^{\pm}(\sigma_0)=k_x^2+k_y^2\pm\gamma_0 k_xk_y+k_z^2\big(1\pm\frac{v\sigma_0^2}{2}\big)$. 
We should emphasise now that the unusual coupling between the order parameter, $\sigma_0$, and the 
frequencies, $k_z=\omega/c_0$, in $D_{\vec{k}}^{\pm}(\sigma_0)$ will be responsible for the 
first order character of the quantum phase transition.

Eqs. (\ref{Eqs-SP}) determine the phase diagram of the model.
By solving the above set of equations self consistently we obtain the behaviour depicted in 
Fig.\ \ref{Fig-Magnetization}.
We observe that 
while for the Heisenberg model the OP goes smoothly to zero at $g_c$ (indicating
a 2nd order PT), frustration brings the system closer to the QCP and the OP jumps 
discontinuously to zero at $g_c$, indicating a 1st order PT. The same is true for the
spin gap (see the inset) when the transition is approached from the
nonmagnetic side. 

%
%**********************************************************************************************
\begin{figure}[h]
\includegraphics[scale=0.4]{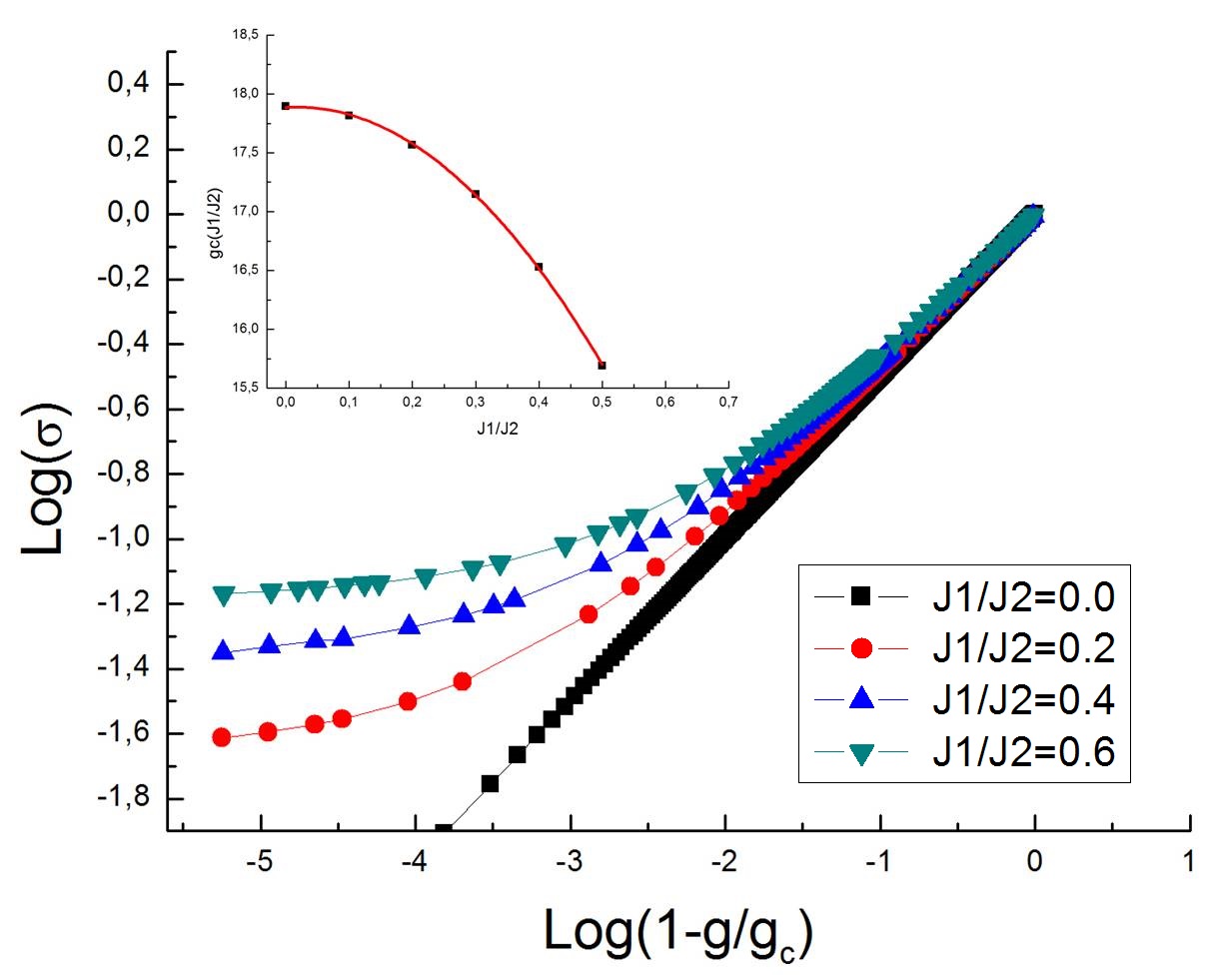}
\caption{Plot of $Log(\sigma)\times Log(1-g/g_c)$ for different values of $J_1/J_2$. A 2nd order 
PT, at the mean field (large $N$) level, produces a straight line (black squares) with slope  
$\beta=1/2$, where $\sigma\rightarrow 0$ as $g\rightarrow g_c$, as expected for $J_1=0$
(Heisenberg model). With frustration, $J_1\neq 0$ (red, blue and green symbols), however, 
$\sigma\neq 0$ as $g\rightarrow g_c$ (with increasing saturation value for $\sigma$ with 
increasing $J_1/J_2$)  indicating a 1st order PT.}
\label{Fig-Critical-Regime}
\end{figure}
%**********************************************************************************************
%

To further establish the 1st order nature of the PT, we exhibit, in Fig.\ \ref{Fig-Critical-Regime}, 
the dependence of $Log(\sigma_0)$ as a function
of $Log(1-g/g_c)$. For $J_1=0$ the behaviour is of a straight line with slope given by $\beta=1/2$, 
as expected for a mean field behavior (large $N$) in a 2nd order QPT 
(with the order parameter vanishing continuously at the quantum critical point). For $J_1\neq 0$,
however, we observe that, although away from the critical point the deviation from mean field 
behaviour is very small, closer to $g_c$ the deviation is significant and characteristic of a first 
order quantum phase transition, with $\sigma_0$ saturating as $g\rightarrow g_c$. 

Let us now provide definitive analytical evidence that the transition is indeed
1st order and not a sharp 2nd order PT. We note that the parameter 
values obtained from the self-consistent equations are such that 
we can write $m_0/c_0$ as a function of $\sigma_0$ \cite{Next}
\begin{eqnarray}
\Big(\frac{m_0}{c_0}\Big)^2(\sigma_0)=\frac{\frac{gv}{2}b_1(\sigma_0)}{1-\frac{gv}{2}b_2(\sigma_0)},
\end{eqnarray}
where
$b_1(\sigma_0)=\int\frac{d^3k}{(2\pi)^3}k_z^2\Bigg(\frac{e^{-D_{\vec{k}}^{-}(\sigma_0)/\Lambda^2}}
{D_{\vec{k}}^{-}(\sigma_0)}-\frac{e^{-D_{\vec{k}}^{+}(\sigma_0)/\Lambda^2}}{D_{\vec{k}}^{+}(\sigma_0)}\Bigg)$,
and
$b_2(\sigma_0)=-\int\frac{d^3k}{(2\pi)^3}k_z^2\Bigg( \frac{e^{-D_{\vec{k}}^{-}(\sigma_0)/\Lambda^2}}
{(D_{\vec{k}}^{-}(\sigma_0))^2}-\frac{e^{-D_{\vec{k}}^{+}(\sigma_0)/\Lambda^2}}{(D_{\vec{k}}^{+}(\sigma_0))^2}\Bigg)$,
and within such approximation we can rewrite the system of self-consistent equations
(\ref{Eqs-SP}) in terms of a single self-consistent variable, namely
\begin{eqnarray}
\sigma_0=\sqrt{f(\sigma_0,m_0(\sigma_0))}.
\label{Eq-Single-SP-Eq}
\end{eqnarray} 
%

%**********************************************************************************************
\begin{figure}[t]
\includegraphics[scale=0.36]{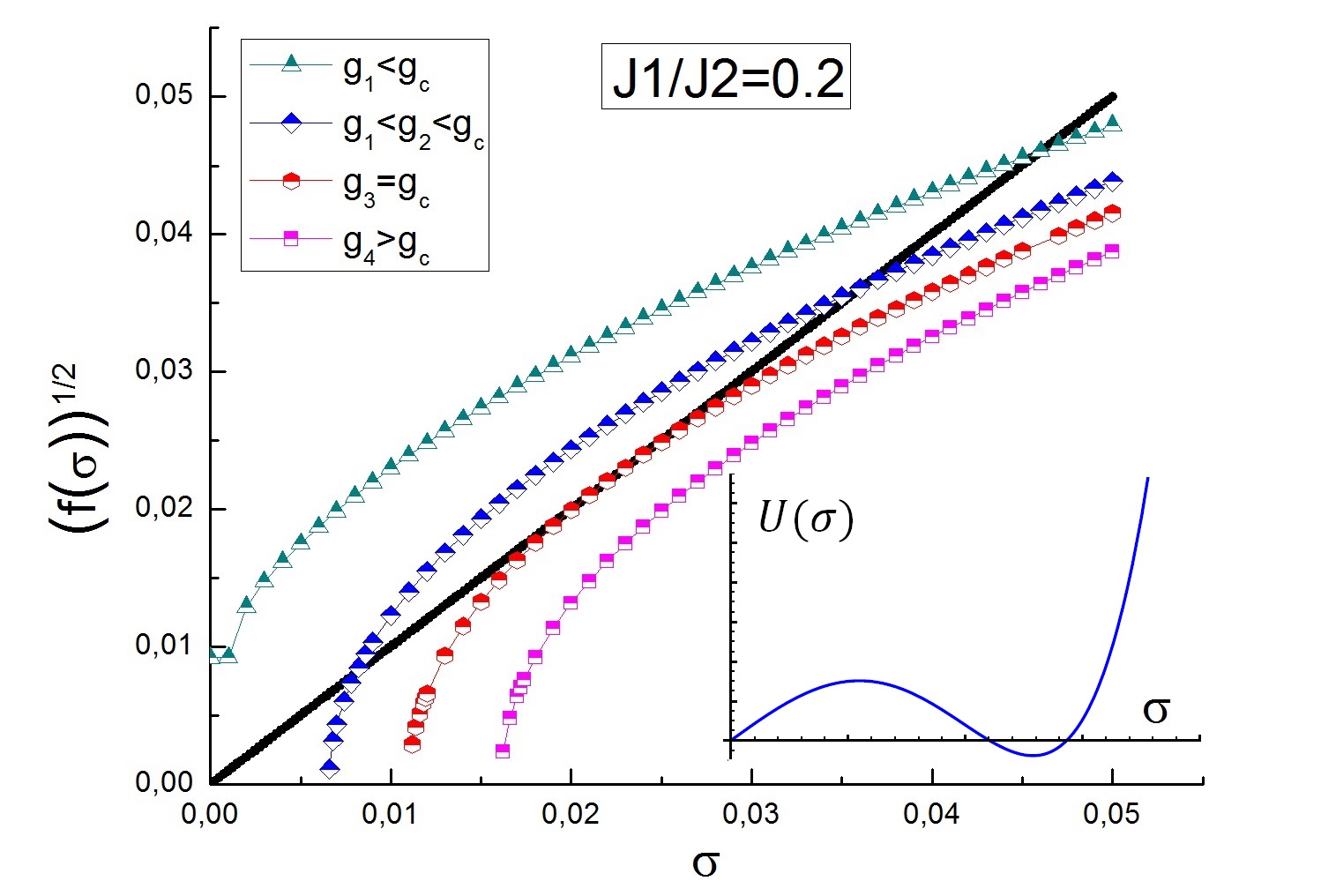}
\caption{Solution of Eq. (\ref{Eq-Single-SP-Eq}) for fixed $J_1/J_2=0.2$. For $g<g_c$ (green triangles and blue diamonds) there is
only one stable equilibrium solution for $\sigma$ (inset: minimum of $U(\sigma)$).
For $g=g_c$ (red circles), Eq. (\ref{Eq-Single-SP-Eq}) produces a nonzero 
value of $\sigma$, while for $g>g_c$ (pink squares) no solution is found and $\sigma$ 
jumps to zero discontinuously above $g_c$, as in a 1st order PT.}
\label{Fig-Intersection}
\end{figure}
%**********************************************************************************************

Fig. \ref{Fig-Intersection} shows the plots of Eq. (\ref{Eq-Single-SP-Eq}) for $J_1/J_2=0.2$ and for
different values of the coupling constant $g$. For $g=g_1<g_c$ (green triangles) $y_1(\sigma_0)=\sqrt{f(\sigma_0,m_0(\sigma_0))}$ 
crosses the straight line $y_2(\sigma_0)=\sigma_0$ at only one point, giving the value of the staggered
magnetisation for this value of the coupling constant. For $g_1<g=g_2<g_c$ (blue diamonds) however
we see that $y_1(\sigma_0)$ and $y_2(\sigma_0)$ cross twice. The first (smaller) value of the magnetisation,
however, corresponds to a local maximum of the free energy $U(\sigma)$ (inset: unstable fixed point) and shall be discarded, 
while the magnetisation is then determined by the second (higher) crossing point solely. By further 
increasing the coupling constant $g=g_3=g_c$ (red circles) we find a single critical solution 
to Eq. (\ref{Eq-Single-SP-Eq}) giving a finite, nonzero and sizeable value for the staggered 
magnetization, which, however, ceasses to exist for $g=g_4>g_c$ 
(pink squares). The fact that the sublattice magnetisation jumps to zero discontinuously for $g>g_c$
indicates the 1st order nature of the QPT.

%**********************************************************************************************
\begin{figure}[t]
\includegraphics[scale=0.38]{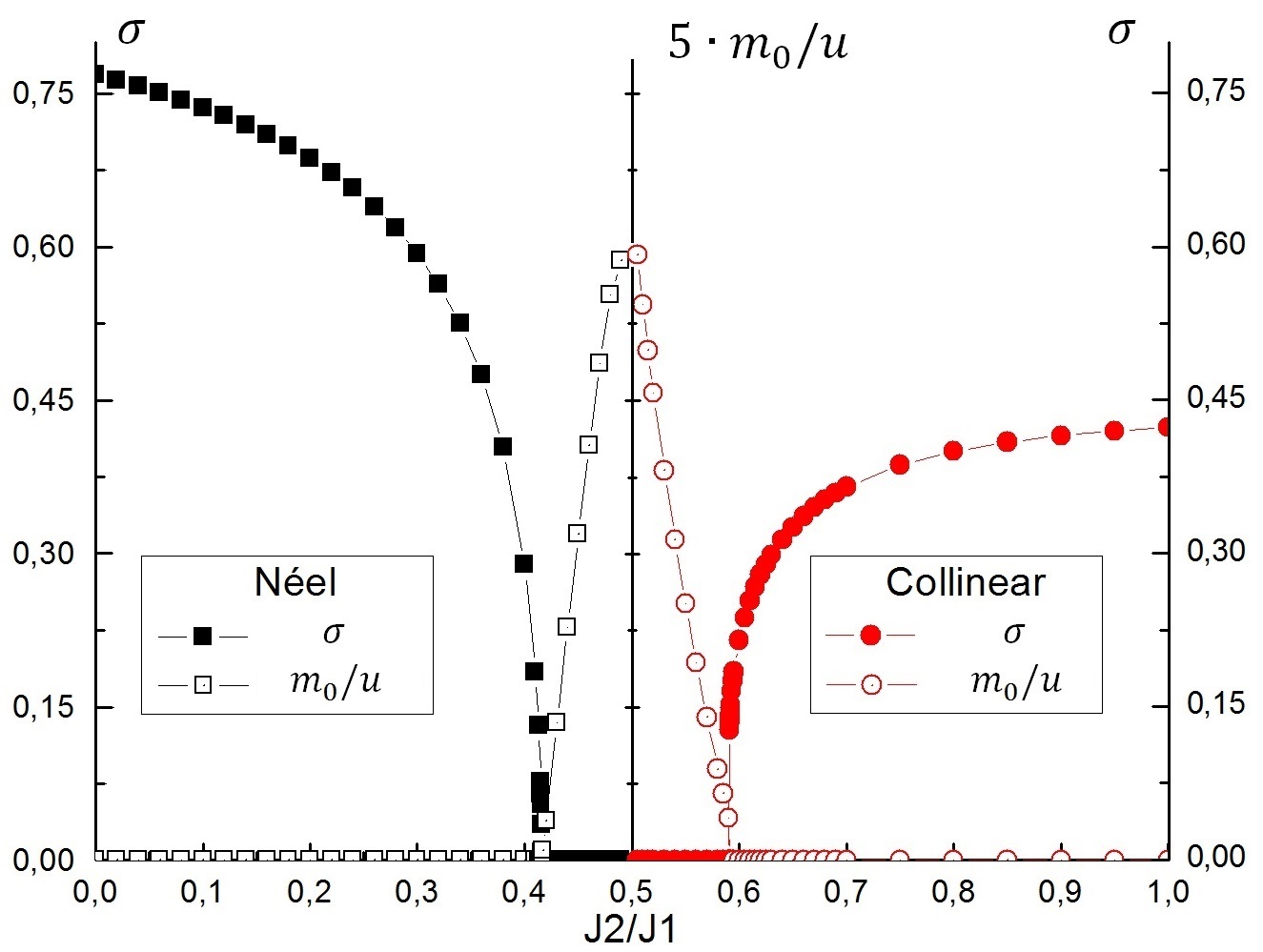}
\caption{Complete phase diagram of the $J_1-J_2$ model
generated by the NLSM for both N\' eel and collinear orders,
$0\leq\alpha\leq 1$. The N\' eel order parameter $\sigma$ 
(filled black squares) vanishes continuously indicating a 2nd 
order PT while the collinear order parameter $\sigma$ (filled 
red circles) jumps to zero at the critical point, indicating a 1st 
order PT. The spin gaps $m_0/u$ (empty black squares
and empty red circles, scaled up by a factor $5$ for clarity
and $u=\sqrt{2}SaJ_1$) cross linearly at the classical border at $\alpha=0.5$, in 
agreement with DMRG \cite{DMRG}.}
\label{Fig-Phase-Diag}
\end{figure}
%**********************************************************************************************

It is important to emphasise that the magnon dispersion along the
collinear directions $k_x=k_y=k$ is given by 
$\omega_{NLSM}(\vec{k})=c_{-}\sqrt{2-\gamma_0}|k|$, 
with $c_-^{-2}=c_0^{-2}-c_1^{-2}/2$ \cite{Conceicao,Dai}, 
and thus acquires an 
imaginary part beyond the border at $\alpha<0.5$ when $\gamma_0>2$, 
showing that the magnetic 
excitations of the collinear state move from $q=(\pi,0)$ and/or $q=(0,\pi)$
towards the one of the N\' eel state at $q=(\pi,\pi)$, as expected. The
complete phase diagram obtained within the NLSM formalism, for
the whole range $0\leq\alpha\leq 1$ is given in Fig.~\ref{Fig-Phase-Diag}.

We have obtained the complete phase diagram of the antiferromagnetic $J_{1}$-$J_{2}$ 
Heisenberg model within the framework of the $O(N)$ nonlinear 
sigma model. We have found that the two magnetically ordered phases, N\' eel order
for $\alpha \leq 0.4$, and collinear order for $\alpha\geq 0.6$, are separated by 
a nonmagnetic region at $0.4\leq \alpha \leq 0.6$ where a gapped spin liquid 
is found. The transition at $\alpha=0.4$ is of the second order while the one at 
$\alpha=0.6$ is of the first order and the spin gaps cross linearly at $\alpha=0.5$. Our 
results are exact at $N\rightarrow\infty$ and agree with numerical results 
from different methods.   

This work was  supported by CNPq and FAPERJ.

\end{document}